\documentclass[twocolumn,preprintnumbers,amsmath,aps]{revtex4}
\usepackage{graphicx}
\usepackage{dcolumn}
\usepackage{bm}
\usepackage{subfigure}
\usepackage{color}
\begin{document}
%%%%%%%%%%%%%%%%%%%%%%%%%%%%
\newcommand{\kvec}{\mbox{{\scriptsize {\bf k}}}}
%%%%%%%%%%%%%%%%%%%%%%%%%%%%
\def\eq#1{(\ref{#1})}
\def\fig#1{figure\hspace{1mm}\ref{#1}}
\def\tab#1{table\hspace{1mm}\ref{#1}}
%%%%%%%%%%%%%%%%%%%%%%%%%%%%
\title{The electronic band structures and optical absorption spectra for incommensurate twisted few-layers graphene}
\author{D. Ghader$^{1}$}\email{Doried.Ghader@univ-lemans.fr}
\author{A. Khater$^{1}$}\email{antoine.khater@univ-lemans.fr}
\author{D. Szcz{\c e}{\' s}niak$^{2}$}
%%%%%%%%%%%%
\affiliation{$^{1}$IMMM UMR 6283 CNRS, LUNAM, University du Maine, 72000 Le Mans, France}
\affiliation{$^{2}$Qatar Environment and Energy Research Institute, Qatar Foundation, PO Box 5825, Doha, Qatar}
%%%%%%%%%%%%
\date{\today}
\begin{abstract}
%%%%%%%%%%%%%%%%%%%%%%%%%%%%%%%%%%%%%%%%%%%%%%%%%%%

A theoretical model is presented to compute the electronic band structures and optical absorption spectra for twisted incommensurate few-layers graphene (tLFG) systems of arbitrary architecture. This is accomplished using an integrated virtual crystal approximation and tight-binding (VCA-TB) method, where the VCA is achieved by a mathematical averaging formalism developed over the quasi-infinite ensemble of interlayer bond configurations. The results show that the low-energy electronic band structures of the incommensurate tLFG systems of N graphene layers are formed of N overlapping Dirac cones centered on the K-points of the Brillouin zones of the graphene layers. This yields effective gaps between the saddle points of the valence and conduction bands of these tFLG systems, that are tunable with the incommensurate twist angles. The optical absorption spectra are also  calculated for these systems; in particular, they display clear spectral peaks which correspond to the accessible electronic transitions across the gaps between the valence and conduction bands. The optical absorption results highlight the potential of these incommensurate tFLG systems for applications in graphene based tandem cells.

%%%%%%%%%%%%%%%%%%%%%%%%%%%%%%%%%%%%%%%%%%%%%%%%%%%
\end{abstract}
\maketitle
\section{Introduction}

Bilayer graphene may be stacked in three different configurations, namely the AB-type (Bernal stacking), AA-type and twisted-type. The twisted bilayer graphene (tBLG) is the most general, consisting of two sheets of graphene rotated by an arbitrary twist angle $\theta$ other than an integer multiples of $60^{\circ}$. Each of the three stacking types exhibits a different low energy electronic structure, and hence a different optical response \cite{zhang1}-\cite{bao}. The interlayer interaction in tBLG, in particular, modifies significantly  the low-energy band structure and yields novel electronic features not present in monolayer graphene or other bilayer graphenes \cite{lopes1}-\cite{ghader}. The tBLG, therefore, has the richest optical absorption spectrum, owing to its complex low energy band structure, where the optical response is observed to be highly tunable through the twist angle $\theta$ \cite{moon1},\cite{liang1}.

Commensuration in tBLG occurs at very specific twist angles \cite{mele1}. As a consequence, experimentally prepared tBLG structures are in general incommensurate. The electronic band structures of bilayer and few-layers graphene have been investigated intensively using absorption measurement  experiments  \cite{zhang2}-\cite{mak}. Theoretically, various models have been developed to calculate the electronic structure in commensurate tBLG \cite{lopes1},\cite{shallcross},\cite{bistritzer1}, \cite{mele1},\cite{bistritzer2}-\cite{uchida}.  These models were also used to calculate the optical absorption for  tBLG for a wide range of rotation angles, doping conditions, and interlayer potential asymmetry  \cite{moon1},\cite{wang2},  \cite{moon2}-\cite{tabert}. However, the optical properties of incommensurate twisted bilayer (tBLG) and few-layers graphene (tFLG) have not been investigated yet on the basis of a systematic theory for incommensurate heterostructures.

Incommensurate tBLG constitutes a disordered heterostructure characterized by a quasi-continuum set of interlayer bonds, and a quasi-infinite set of configurations for carbon atoms at the bilayer interface. In a previous work \cite{ghader}, we have developed a non-trivial Virtual Crystal Approximation - Tight Binding (VCA-TB) theory to study the electronic properties in such systems. Our VCA-TB theoretical approach determines the VCA effective medium for the incommensurate tBLG system, derived in a direct manner from the quasi-infinite set of structural configurations at the bilayer interface. The theory consequently determines the VCA-TB Hamiltonian for the effective VCA system and yields the tBLG’s electronic structure and density of states for any incommensurate twist angle $\theta$.

The purpose of the present work is to extend the VCA-TB theoretical approach to study the electronic structure and the optical absorption properties in incommensurate twisted few-layers graphene (tFLG) with arbitrary stacking geometries. The band structure for incommensurate tFLG calculated by the VCA-TB method is found to be formed of overlapped Dirac cones, equal in number to that of the graphene sheets forming the incommensurate heterostructure. The overlap between these Dirac cones gives rise to numerous saddle points in the low-energy band structure and consequently to van Hove singularities (VHS) in the density of states (DOS). The optical absorption spectrum, calculated in the framework of the VCA-TB theory, is found to be characterized by a series of peaks associated with the angle dependent van Hove singularities. The transition energies corresponding to optical absorption peaks are hence directly tunable via the rotation angles defining the twists in the stacking configurations of the incommensurate tFLG.

The rest of this paper is organized as follows. In Section II we extend the VCA-TB model and apply it to determine the band structures and densities of states for incommensurate twisted tFLG. In Section III, we calculate the optical absorption for tFLG for various incommensurate architectures and analyze the results in view of the characteristic features of the band structure. The conclusions are discussed in Section IV.

%%%%%%%%%%%%%%%%%%%%%%%%%%%%%%%%%%%%%%%%%%%%%%%%%%%
\section{Band Structure for Incommensurate Twisted Few-Layers Graphene}
%%%%%%%%%%%%%%%%%%%%%%%%%%%%%%%%%%%%%%%%%%%%%%%%%%%

Consider an incommensurate heterostructure of N twisted graphene layers stacked on top of each other.  The layers are labeled by $n$=1, 2,.., N. Any layer $n+1$ is assumed twisted with respect to adjacent layer $n$ by an incommensurate twist angle $\theta_{n,n+1}$. The $A$ and $B$ sublattices in the graphene layer $n$ are denoted by $A_{n}$ and $B_{n}$, and the corresponding atoms are labeled as $A_{n} (s)$ and $B_{n} (s)$; Fig.1  shows only the first eighteen sites of the infinite set of sites per layer.
\begin{figure}[!h]
\centering
\includegraphics[scale=0.7]{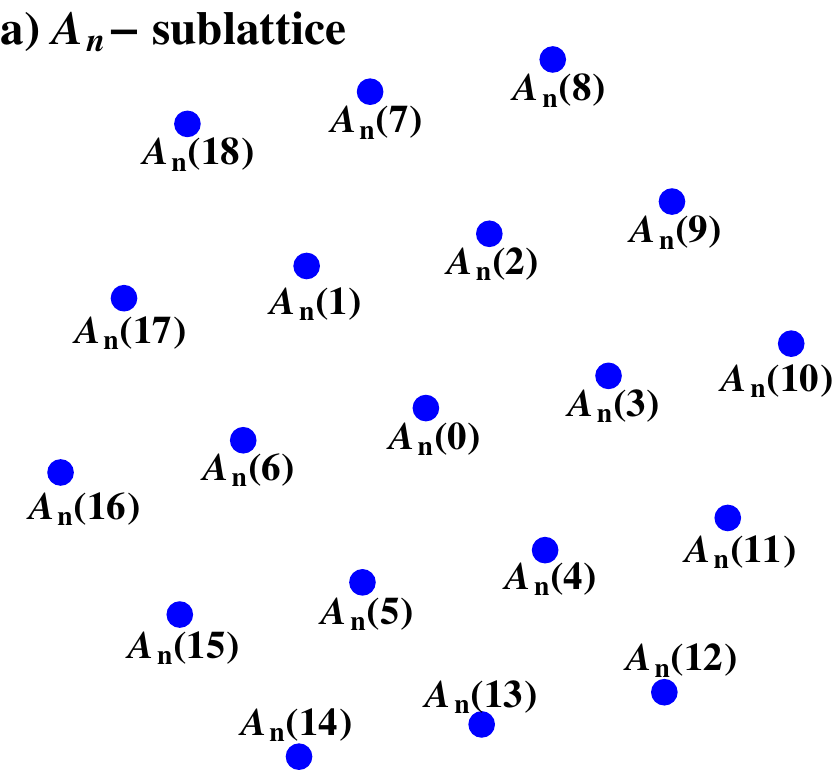}\vspace{0.5cm}
\includegraphics[scale=0.7]{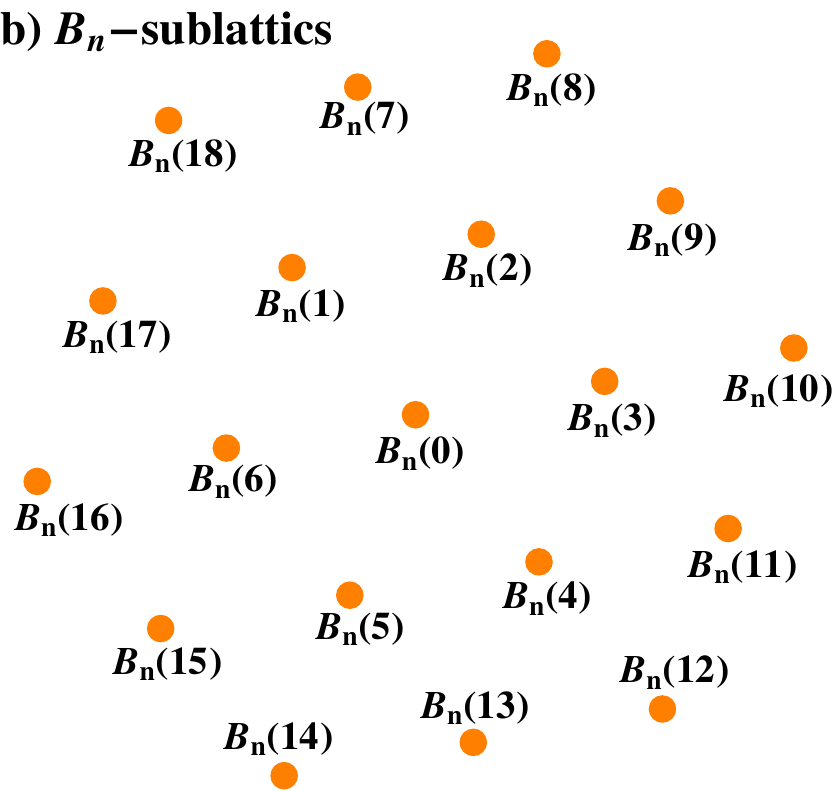}
\caption{Figs. a) and b) present schematic representations of the $A_{n}$ and $B_{n}$ sublattices in layer $n$ of the incommensurate tFLG systems, with representative sublattice atomic sites.}
\end{figure}

In the VCA-TB model, we consider a referential couple of sites in each graphene layer $n$, denoted by ${A_{n} (0),B_{n} (0) }$. The incommensurate tFLG is hence represented by a basis of $2N$ irreducible atomic sites, namely ${A_{1}(0),B_{1}(0),.., A_{N}(0),B_{N}(0)}$. In the VCA-TB method, the effective VCA medium is derived as a direct system average from the quasi-infinite set of structural configurations at bilayer interfaces within the tFLG. The effective medium is hence periodic and one may write the total electronic wave function for the tFLG as
\begin{eqnarray}
\nonumber
\Psi_{\bf k}({\bf r})&=&\sum_{n=1}^{N}C_{A_{n}} \Psi_{{\bf k}; A_{n}} ({\bf r}) +\sum_{n=1}^{N} C_{B_{n}} \Psi_{{\bf k}; B_{n}} ({\bf r}).
\end{eqnarray}
\noindent In our formalism, we consider a single $p_{z}$-orbital per site. The wave function $\Psi_{{\bf k}; \alpha}({\bf r})=1/\sqrt{N^{'}} \sum_{{\bf r}_\alpha}e^{i {\bf k}.{\bf r}_{\alpha}}\Phi_{\alpha}({\bf r}-{\bf r}_{\alpha})$, $\alpha=A_{n}$ or $B_{n}$, is the VCA-TB Bloch wave function from the $\alpha$ sublattice in layer $n$. $\Phi_{\alpha}$ denotes a $2p_{z}$ atomic orbital and the letter $N^{'}$ is the number of unit cells in the crystal. ${\bf r}_{\alpha}$ is the position of an $\alpha$ type atom with respect to a chosen origin. The coefficient $C_{\alpha}$ ($\alpha=A_{n}$, or $B_{n}$) is the $\alpha$ sublattice contribution to the total wave function and {\bf k} is the wave vector.
The band structure in the incommensurate tFLG system may be determined by solving the secular equation
\begin{eqnarray}
\det[\bar{H}-E({\bf k})S]=0,
\end{eqnarray}
\noindent or equivalently by determining the eigen-energies $E({\bf k})$ of the matrix $S^{-1} \bar{H}$. The matrices $\bar{H}$ and $S$ respectively denote the VCA-TB Hamiltonian and overlap matrices with matrix entities  $S_{\alpha \beta}= \left< \Psi_{{\bf k}; \alpha} | \Psi_{{\bf k}; \beta} \right> = \iiint {\rm d}^{3} {\bf r} \Psi_{{\bf k}; \alpha}^{*} ({\bf r}) \Psi_{{\bf k}; \beta} ({\bf r})$ and $\bar{H}_{\alpha \beta}= \left< \Psi_{{\bf k}; \alpha} | \hat{\bar{H}} | \Psi_{{\bf k}; \beta} \right> = \iiint {\rm d}^{3} {\bf r} \Psi_{{\bf k}; \alpha}^{*} ({\bf r}) \hat{\bar{H}} \Psi_{{\bf k}; \beta}({\bf r})$; $\alpha, \beta \in \{ A_{n}, B_{n};n=1,...,N \}$. 
\indent The in-plane entities of the VCA-TB Hamiltonian matrix $\bar{H}$ may be readily written as \cite{kundu}
\begin{eqnarray}
\nonumber
\bar{H}_{A_{n} A_{n}}=\bar{H}_{B_{n} B_{n}}&=&E_{2p} \left[1 + 2s_{1} \sum_{i=1}^{3} {\rm cos}({\bf a}_{i}(n).{\bf k}) \right]
\nonumber\\&+& 2 \gamma_{1} \sum_{i=1}^{3} {\rm cos}({\bf a}_{i}(n).{\bf k}),\nonumber
\end{eqnarray}
\begin{eqnarray}
\nonumber
\bar{H}_{A_{n} B_{n}}&=&\bar{H}_{B_{n} A_{n}}^{*} \nonumber \\
&=&E_{2p} \left\{ s_{0} \gamma_{\bf k}(n) + s_{2} \left[ 2 {\rm cos} ({\bf a}_{1}(n).{\bf k}) + e^{i {\bf a}_{4}(n).{\bf k}}\right]\right\} \nonumber \\
&+&\gamma_{0} \gamma_{\bf k}(n) + \gamma_{2} \left[ 2 {\rm cos} ({\bf a}_{1}(n).{\bf k}) + e^{i {\bf a}_{4}(n).{\bf k}}\right].\nonumber
\end{eqnarray}
Here we have defined the lattice vectors $ {\bf a}_{i} (n)$ for $i$=1, 2, 3, 4 for the graphene layer $n$ as ${\bf a}_{i} (n)=R \left(\theta_{1,2}+\theta_{2,3}+⋯+\theta_{n-1,n} \right){\bf a}_{i}(1)$, where $R$ denotes the $2 \times 2$ rotation matrix, and the vectors ${\bf a}_{1} (1)=a(\sqrt{3},0)$, ${\bf a}_{2} (1)=\frac{\sqrt{3}}{2} a(1,\sqrt{3})$, ${\bf a}_{3} (1)={\bf a}_{2} (1)-{\bf a}_{1} (1)$, ${\bf a}_{4} (1)=a(0,2)$.
The parameters ${\gamma_{0},\gamma_{1},\gamma_{2}}$ and  ${s_{0},s_{1},s_{2}}$ denote the in-plane first, second and third nearest neighbor hopping and overlap integrals, respectively. $E_{2p}$ denotes the onsite energy and $\gamma_{\bf k} (n)=1+e^{i {\bf a}_{1} (n).{\bf k}}+e^{i {\bf a}_{2} (n).{\bf k}}$.\\
\indent Due to the incommensurate nature of the tFLG stacking, the carbon atoms in these systems are characterized by a quasi-infinite ensemble of structural configurations with a quasi-continuum set of interlayer bond lengths and orientations. The key idea in the VCA-TB approach is to determine the ensemble average for the interlayer hopping integrals between neigboring sites, over the quasi-infinite ensemble of structural configurations. We use the environment-dependent tight binding approach, \cite{landgraf}, \cite{tang}, to compute the values of the interlayer hopping integrals and of their upper limit as a function of the randomly distributed bond lengths and configurations. The VCA is achieved by a mathematical averaging formalism developed over the quasi-infinite ensemble of interlayer bond configurations \cite{ghader2}. This model approach has been applied successfully for the tBLG systems with arbitrary incommensurate twist angles \cite{ghader}. We compute in the present work for the generalized tFLG systems, the ensemble averages of the interlayer elements of the VCA-TB Hamiltonian matrix $\bar{H}$ as
\begin{eqnarray}
\nonumber
\bar{H}_{A_{n} A_{m}}&=& [ \bar{t} \left(A_{n}(0) A_{m}(0) \right)+ \bar{t} \left(A_{n}(0) A_{m}(1) \right)\nonumber \\
&\times& \sum_{s=1}^{6}{\rm cos} \left( \left({\bf r}_{A_{m}(s)}-{\bf r}_{A_{m}(0)} \right).{\bf k} \right)] \nonumber \\
&\times& \left(\delta_{n,m-1} + \delta_{n,m+1} \right).\nonumber
\end{eqnarray}
${\bf r}_{A_{n} (s)}$ denotes the position vector of the $A_{n} (s)$  atom as in Fig.1. The quantity $ \bar{t} \left(A_{n} (0) A_{m} (s) \right)$ denotes the average for the interlayer hopping integral between atoms $A_{n} (0)$ and $A_{m} (s)$, computed using the VCA-TB approach \cite{ghader}. The Kronecker deltas represent the interactions between adjacent graphene layers.

\begin{figure}[!h]
\centering
\includegraphics[width=0.5\columnwidth]{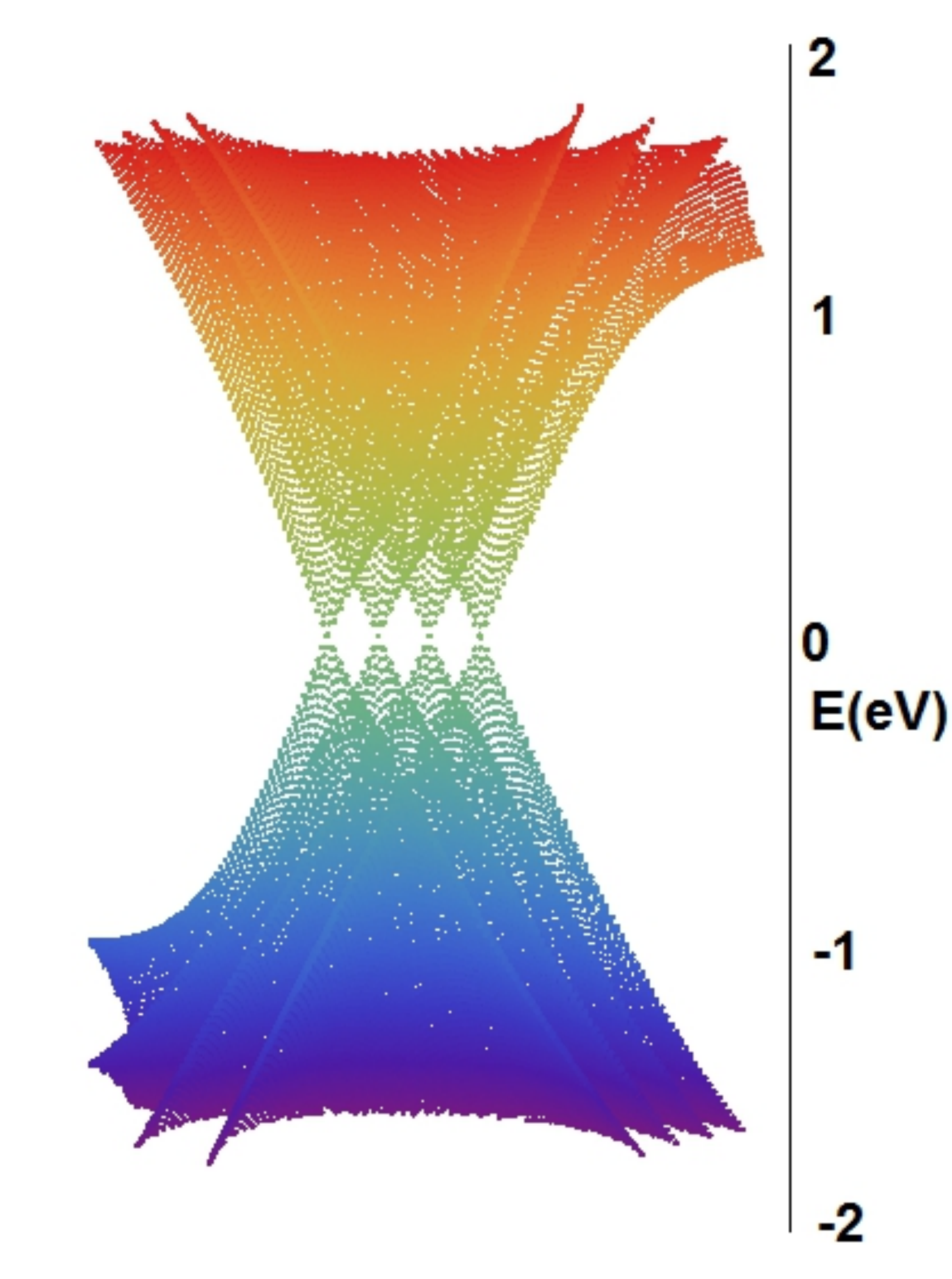}
\caption{The numerically calculated low-energy electronic band structure for the incommensurate tFLG ($N=4$,$\theta_{12}=2^{\circ}$,$\theta_{23}=2^{\circ}$,$\theta_{34}=2^{\circ}$), plotted over the juxtaposed set of the Brillouin zones corresponding to the incommensurably twisted graphene layers forming the graphene heterostructure. Our numerical results show that the band structure of a tFLG composed of $N$ graphene layerss is formed of $N$ overlapped Dirac cones.}
\label{fig2}
\end{figure}

Next we calculate the elements of the overlap $S$ matrix. The interlayer overlap integrals are neglected owing to the large perpendicular distance $d_{\perp}$ between the two graphene layers $\left(d_{\perp} \approx 3.35 {\rm \AA} \right)$. Note in this case that the $S$ matrix eelements in the plane of a graphene layer are invariant under the twist angle. The nonzero matrix elements of the overlap matrix $S$ may be represented as
\begin{equation}
\nonumber
S_{A_{n} A_{n}}=S_{B_{n} B_{n}}= 1 + 2s_{1} \sum_{i=1}^{3} {\rm cos} ({\bf a}_{i}(n).{\bf k}),
\end{equation}
\begin{equation}
\nonumber
S_{A_{n} B_{n}}=S_{B_{n} A_{n}}^{*}= s_{0} \gamma_{\bf k}(n) + s_{2} \left[ 2 {\rm cos} ({\bf a}_{1}(n).{\bf k})+ e^{i {\bf a}_{4}(n).{\bf k}} \right].
\end{equation}
\indent At this stage, one can solve numerically the secular Eq.(1) and determine the band structure for any given tFLG system with any number of graphene sheets $N$, and incommensurate twist angles $\theta_{n,n+1}$. The numerical values for the various averaged interlayer hopping integrals are presented in reference \cite{ghader}, whereas the numerical values for the onsite energy $E_{2p}$, the in-plane hopping integrals {$\gamma_{0}$,$\gamma_{1}$,$\gamma_{2}$}, and the overlap integrals {$s_{0}$,$s_{1}$,$s_{2}$} are taken from reference \cite{kundu}.\\
\indent To illustrate the numerical results, we present in Fig.2 the calculated low energy electronic band structure for the tFLG with four incommensurable graphene sheets ($N=4$,$\theta_{12}=2^{\circ}$,$\theta_{23}=2^{\circ}$,$\theta_{34}=2^{\circ}$); the results demonstrate that the band structure of a tFLG composed of N graphene sheets is formed of $N$ overlapped Dirac cones. This overlap significantly gives rise to numerous saddle points for the valence and conduction bands in the low energy electronic band structure for these systems.\\

\begin{figure}[!h]
\centering
\includegraphics[width=0.47\columnwidth]{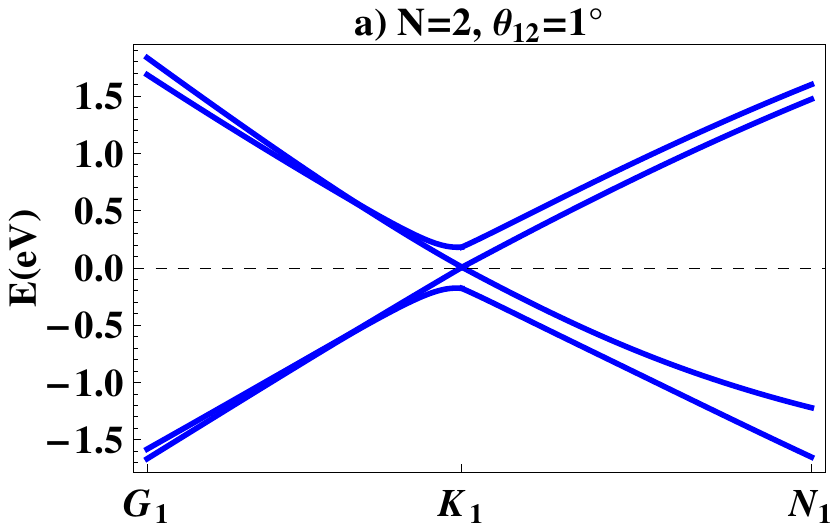}\hspace{0.2cm}
\includegraphics[width=0.47\columnwidth]{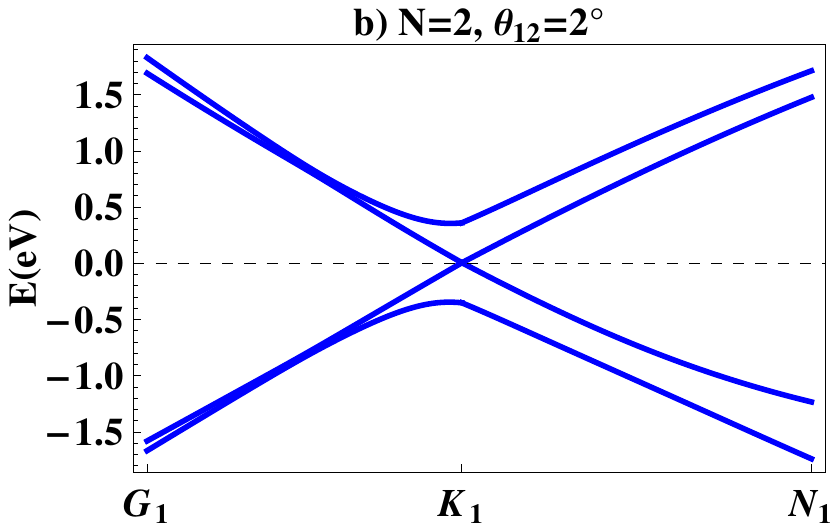}\\ \vspace{0.2cm}
\includegraphics[width=0.47\columnwidth]{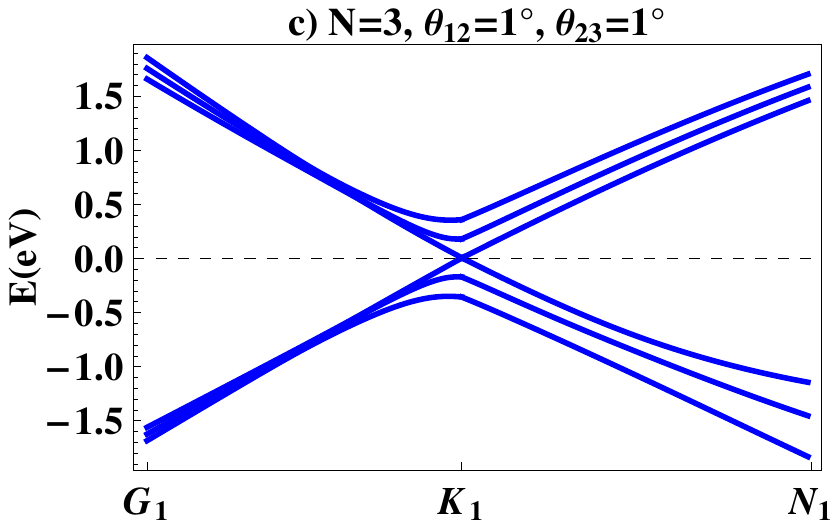}\\ \vspace{0.2cm}
\includegraphics[width=0.47\columnwidth]{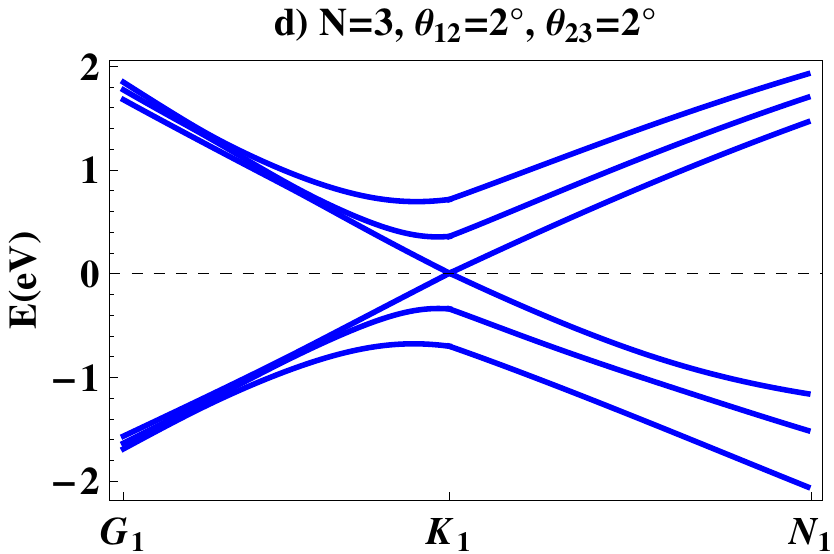}\hspace{0.2cm}
\includegraphics[width=0.47\columnwidth]{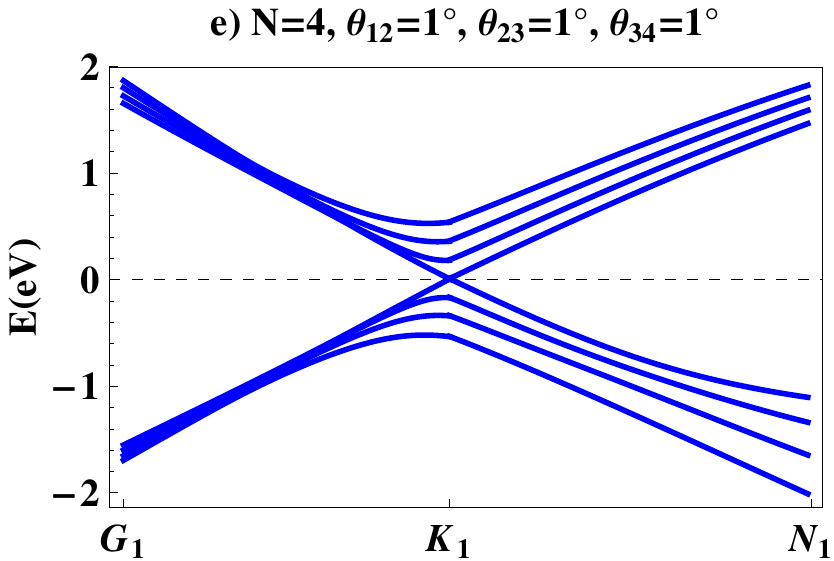}
\caption{Calculated low energy electronic band structures for tBLG and tFLG systems, along the high symmetry axes {$\Gamma K_{1}$ and $K_{1}M_{1}$}. Plotted over the BZ of a reference graphene layer in a given system, these correspond to:  (a) tBLG ($N=2$,$\theta_{12}=1^{\circ}$), (b) tBLG ($N=2$,$\theta_{12}=2^{\circ}$), (c) tFLG ($N=3$,$\theta_{12}=1^{\circ}$,$\theta_{23}=1^{\circ}$), (d) tFLG ($N=3$,$\theta_{12}=2^{\circ}$,$\theta_{23}=2^{\circ}$), and (e) tFLG ($N=4$,$\theta_{12}=1^{\circ}$,$\theta_{23}=1^{\circ}$,$\theta_{34}=1^{\circ}$). The results show the saddle points of the valence and conduction bands with associated gaps due to the overlapping of the Dirac cones from the system graphene layers.}
\label{fig3}
\end{figure}

\begin{figure*}[ht]
\centering
\includegraphics[width=0.65\columnwidth]{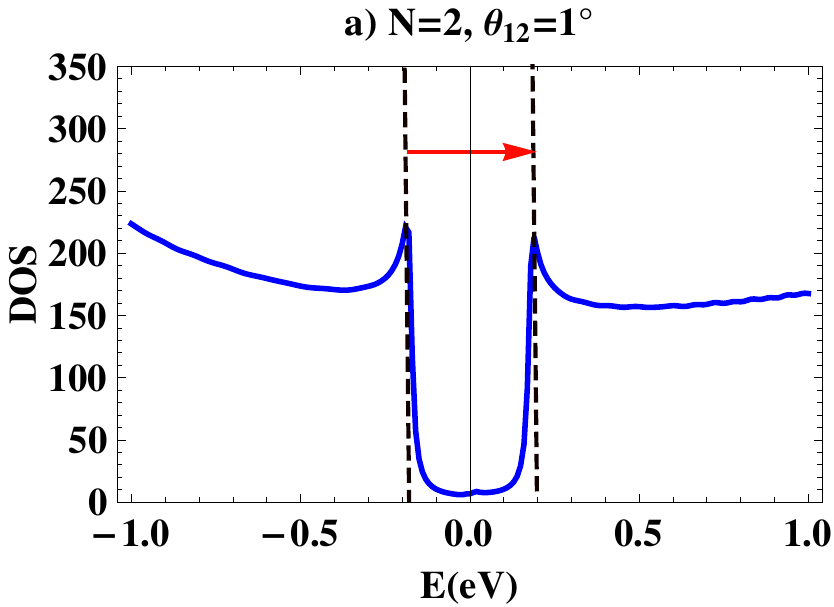}\hspace{0.2cm}
\includegraphics[width=0.65\columnwidth]{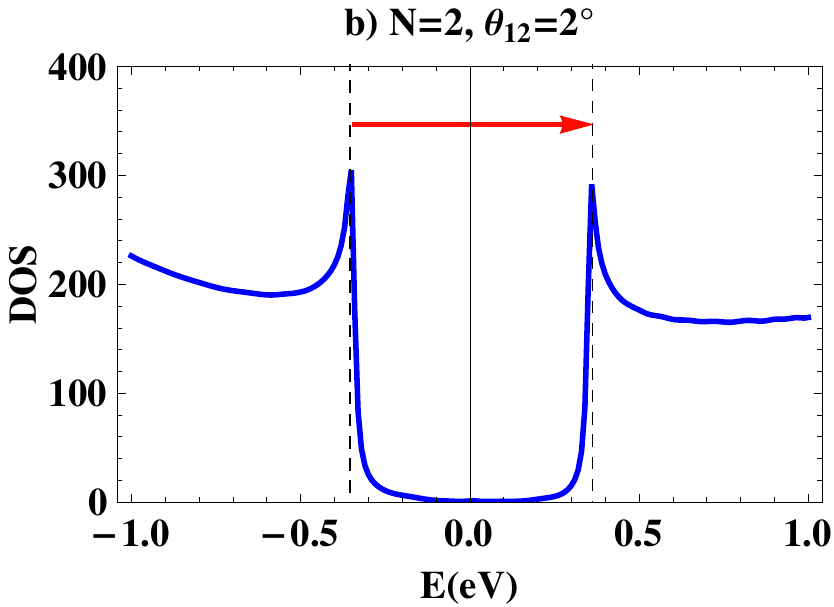}\hspace{0.2cm}
\includegraphics[width=0.65\columnwidth]{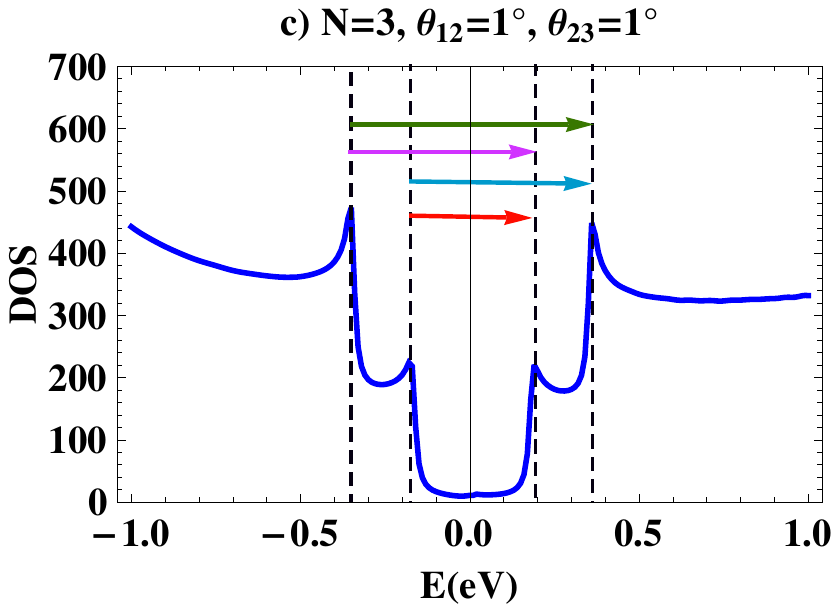}\vspace{0.5cm}
\includegraphics[width=0.65\columnwidth]{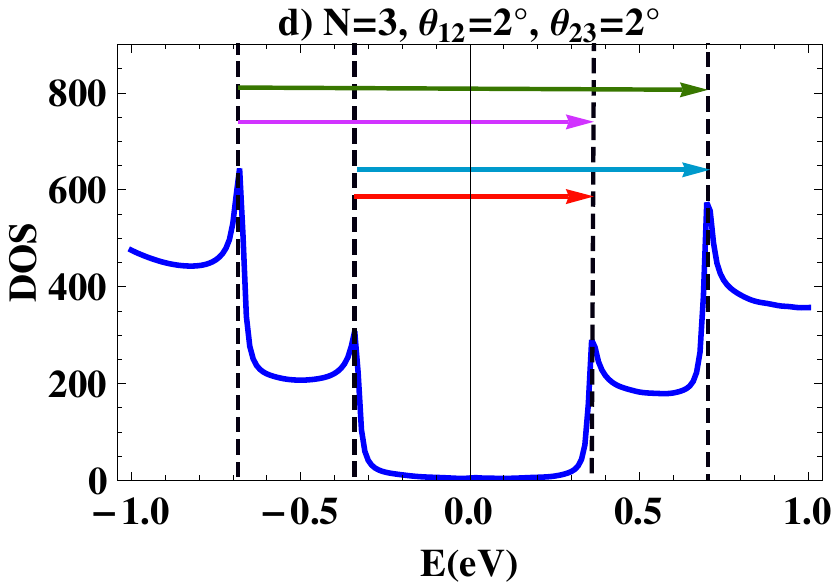}\hspace{0.2cm}
\includegraphics[width=0.65\columnwidth]{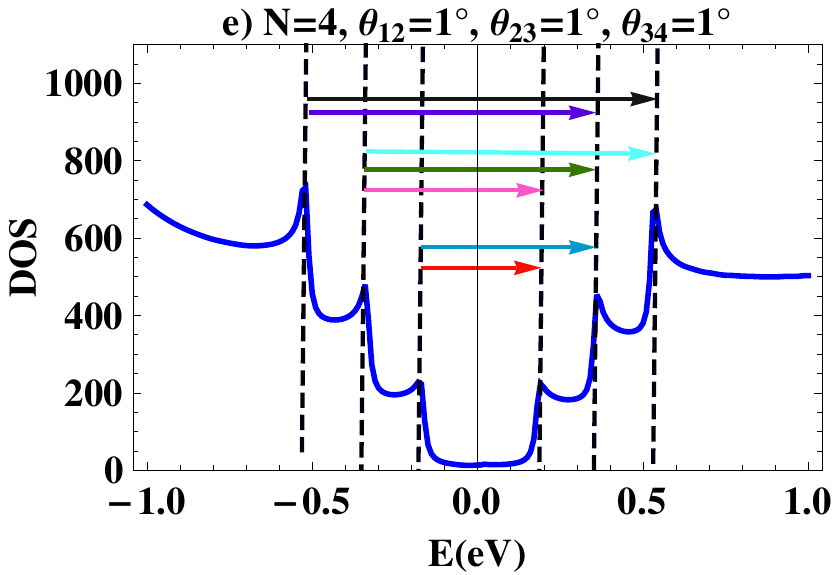}
\caption{Calculated electronic DOS for the tBLG and tFLG systems, detailed in Figs.3 as (a), (b),  (c), (d), and (e), along the high symmetry axes {$\Gamma K_{1}$ and $K_{1}M_{1}$}; the results are plotted over the first BZ of a reference graphene layer.}
\label{fig4}
\end{figure*}

To further illustrate the behavior of the valence and conduction bands, and their saddle points, with the variation of the graphene incommensurate stacking, we present in Figs.3 the low energy electronic band structures for two tBLG and three tFLG systems, along the high symmetry axes {$\Gamma K_{1}$ and $K_{1}M_{1}$ } in the Brillouin zone of layer 1.  To zoom on the band structures near the saddle points, we opt in the BZ without loss of generality for the points $G_{1}$ and $N_{1}$ along  $\Gamma K_{1}$ and $K_{1} M_{1}$ respectively, with $\overrightarrow{\Gamma G_{1}}=0.85 \overrightarrow{\Gamma K_{1}}$ and $ \overrightarrow{K_{1} N_{1}} = 0.33 \overrightarrow{K_{1} M_{1}}$.

In particular, Figs.3a, 3b, and 3b, correspond to the band structures for two tBLG systems with increasing twist angles, namely ($N=2$,$\theta_{12}=1^{\circ}$) and ($N=2$,$\theta_{12}=2^{\circ}$), 
whereas Figs.3c, 3d and 3e, correspond to the band structures for three tFLG systems with increasing twist angles, namely ($N=3$,$\theta_{12}=1^{\circ}$,$\theta_{23}=1^{\circ}$), ($N=3$,$\theta_{12}=2^{\circ}$,$\theta_{23}=2^{\circ}$), and ($N=4$,$\theta_{12}=1^{\circ}$,$\theta_{23}=1^{\circ}$,$\theta_{34}=1^{\circ}$), respectively. Note that the Fermi energy is taken as a base line reference at $E_F=0$.

The saddle points in the low energy electronic  band structures of Figs.3 yield the van Hove singularities in the corresponding densities of states (DOS), calculated along {$\Gamma K_{1}$ and $K_{1}M_{1}$ } and presented in Figs.4. It is clear that the positions of the spectral singularities may be  directly tuned via the incommensurate twist angles of the tFLG systems. Note that the arrows in Figs.4 highlight the electronic  transitions by optical absorption across the gaps between the saddle points of the valence and the conduction bands.

%%%%%%%%%%%%%%%%%%%%%%%%%%%%%%%%%%%%%%%%%%%%%%%%%%%
\section{Optical Absorption}
%%%%%%%%%%%%%%%%%%%%%%%%%%%%%%%%%%%%%%%%%%%%%%%%%%%

The optical absorption of light incident along the normal to the layers of the  tFLG system, is defined as the real part of the dynamic conductivity $\sigma_{xx}$, where the x-axis is in the plane of the layers. The dynamic conductivity may be calculated then along any given direction $\Delta$ in the reciprocal space of the BZ for any of the graphene sheet layers, via the Kubo formula

\begin{eqnarray}
\sigma_{xx}(E)&=& i \frac{g_{s}g_{v}e^{2}}{\omega+i \in } \int_{\Delta}  \frac{dk}{2 \pi }\sum_{\alpha,\beta}\frac{f(E_{\alpha}({\bf k})-f(E_{\beta}({\bf k}))}{E-E_{\beta}({\bf k})+E_{\alpha}({\bf k})+i \in}\nonumber \\
&\times&  \left< \alpha, {\bf k}| \hat{j_{x}} | \beta,{\bf k} \right> \left< \beta, {\bf k}| \hat{j_{x}} | \alpha,{\bf k} \right>.
\end{eqnarray}

\begin{figure*}[ht]
\centering
\includegraphics[width=0.65\columnwidth]{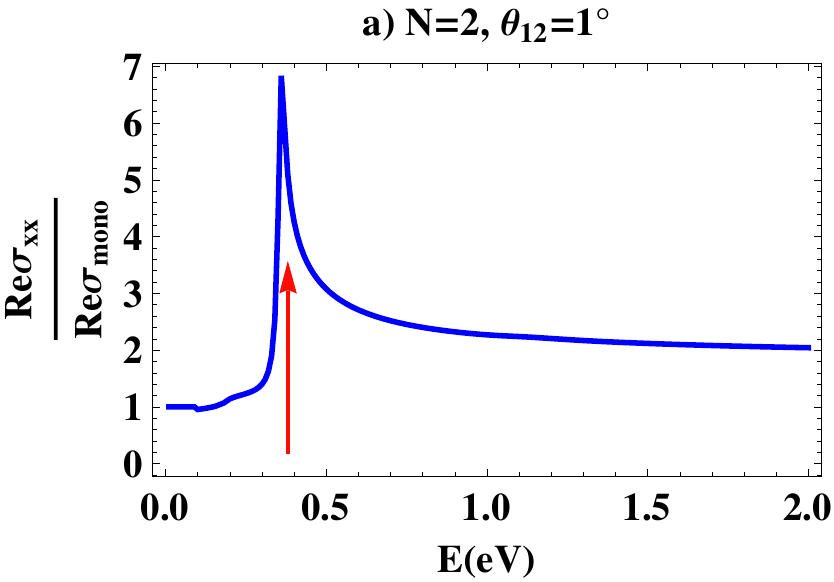}\hspace{0.2cm}
\includegraphics[width=0.65\columnwidth]{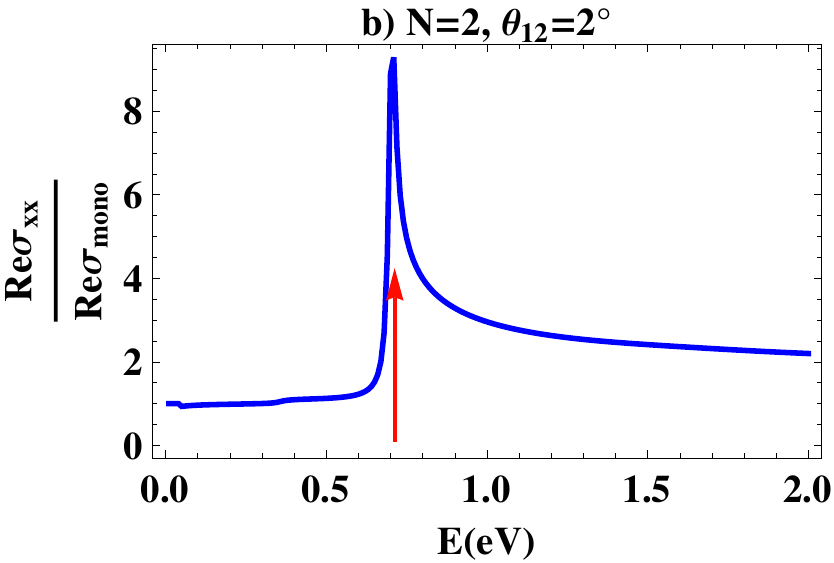}\hspace{0.2cm}
\includegraphics[width=0.65\columnwidth]{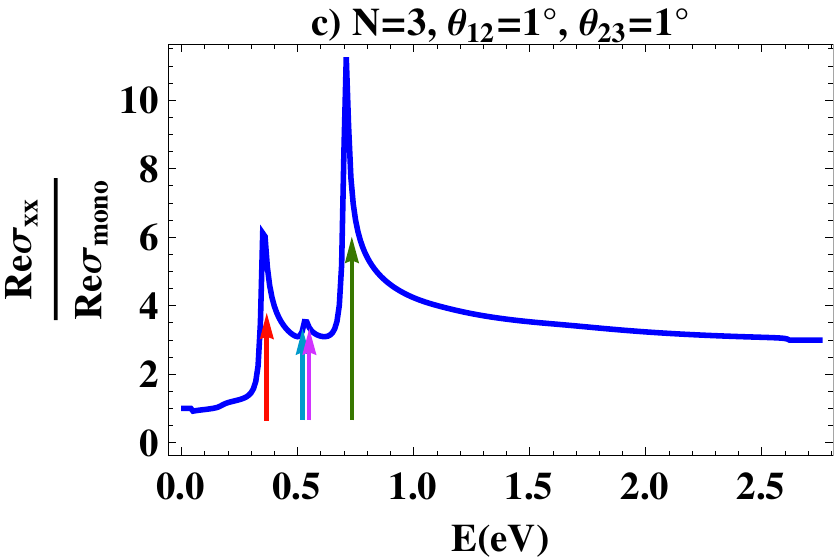}\vspace{0.5cm}
\includegraphics[width=0.65\columnwidth]{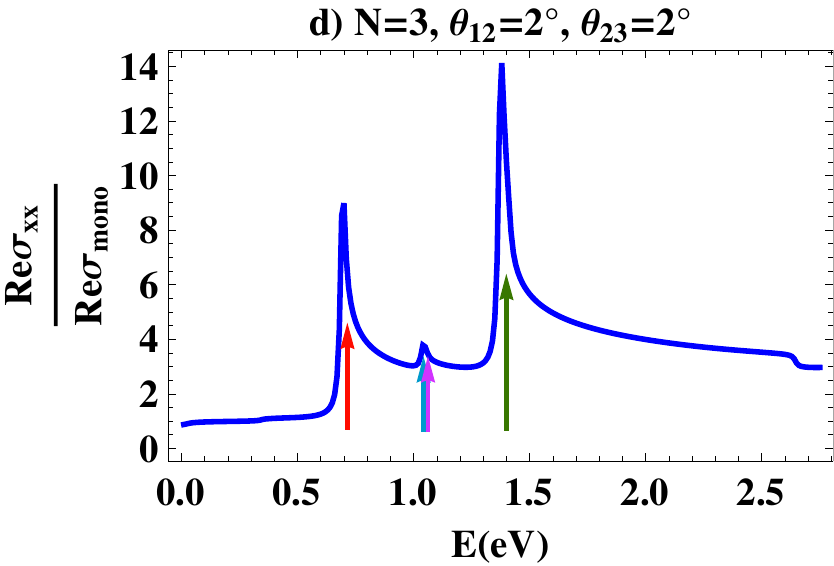}\hspace{0.2cm}
\includegraphics[width=0.65\columnwidth]{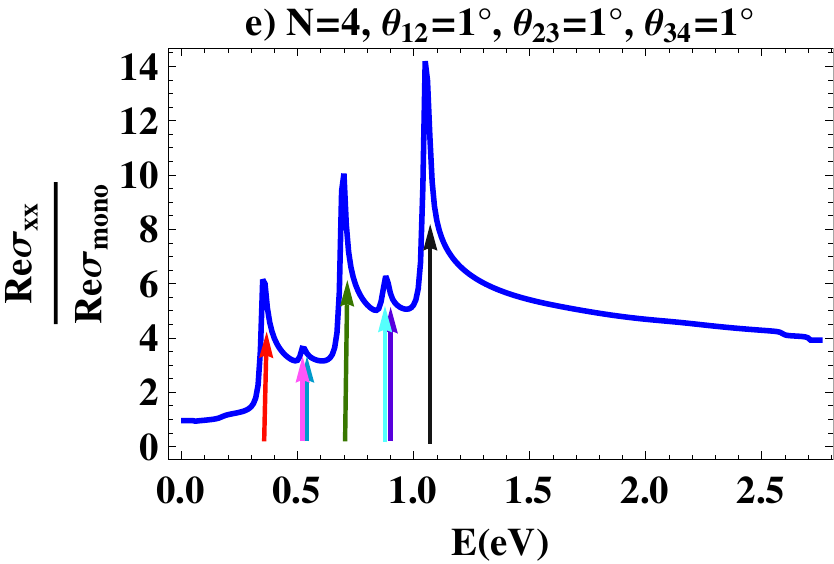}
\caption{Calculated room temperature optical absorption spectra for the tBLG and tFLG systems, (a) and (b), and (c), (d), and (e) respectively, detailed in Figs.3, normalized with respect to the optical absorption spectrum for  monolayer graphene. The absorption spectra are calculated from contributions along the high symmetry axes {$\Gamma K_{1}$ and $K_{1}M_{1}$}.}
\label{fig5}
\end{figure*}

\noindent $\in$ is a positive infinitesimal and $\omega=E/\hbar$. The constants $g_{s}$ and $g_{v}$ denote the spin and electronic valley degeneracies,  respectively ($g_{s}=g_{v}=2$). The operator $\hat{j_{x}}$ is the current operator defined in our formalism as $\hat{j_{x}}=-\partial{\hat{S}^{-1} \hat{\bar{H}}}/\partial{k_{x}}$. The function $f$ is the Fermi function and  $| \alpha,{\bf k}>$ is the $ \alpha$ sub-band eigenstate of the operator $\hat{S}^{-1} \hat{\bar{H}}$ with momentum ${\bf k}$ and eigen-energy $E_{\alpha}({\bf k})$.

\indent Given the calculated electronic states within the VCA-TB approach, we use Eq.(2)  to calculate the dynamic conductivity $\sigma_{xx}(E)$ for the tFLG systems along the high symmetry directions {$\Gamma K_{1}$ and $K_{1}M_{1}$}. This calculation  yields the optical absorption spectra $Re \sigma_{xx}(E)/Re \sigma_{mono}(E)$ for these tFLG systems, normalized with respect to the optical absorption for a graphene monolayer along the same symmetry directions.

\indent In Figs.5, we plot  the normalized optical absorption spectra at room temperature over the low energy intervals of the band structures in the neighborhood of the Fermi level $E_F=0$. The series of results of Figs.5 correspond to the series of the systems of Figs. 3 and 4.

There are a few general remarks which can be made about Figs.5. The normalized optical absorption for the tFLG systems in the neighborhood of the Fermi energy is observed to approach unity. This is a direct consequence of the fact that the energy spectrum of the tFLG systems becomes identical to that of monolayer graphene very close to the $K_{1}$ Dirac point as in Figs. 3. The calculated spectra further display peaks which correspond to the electronic transitions from the saddle points in the valence bands to the saddle points in the conduction bands. The peaks in the optical absorption spectra are highlighted using arrows as in Figs.4, in order to identify each peak with the corresponding transition. Our numerical results demonstrate the possibility to tune the transition energy corresponding to these peaks by the manipulation of the incommensurate twist angles of the layered heterostructures. Finally, the calculated normalized absorption spectra at relatively high energies are observed to converge to a value equal to the number of graphene layers $N$. This result follows from the fact that the electronic structures at such energies are similar to those for monolayer graphene but with a number of modes $N$ times greater than in monolayer graphene.

%%%%%%%%%%%%%%%%%%%%%%%%%%%%%%%%%%%%%%%%%%%%%%%%%%%
\section{Conclusions}
%%%%%%%%%%%%%%%%%%%%%%%%%%%%%%%%%%%%%%%%%%%%%%%%%%%

We have developed a theoretical approach to model and compute  the electronic band structures and optical absorption spectra for twisted incommensurate few-layers graphene tLFG systems of arbitrary architecture. This is accomplished using the integrated VCA-TB method and the environment-dependent tight binding approach, where the VCA is achieved by a mathematical averaging formalism developed over the quasi-infinite ensemble of bond configurations. 

The numerical calculations demonstrate that the electronic band structures of the incommensurate twisted few layers graphene of $N$ sheets, are formed of $N$ overlapping Dirac cones centered on the K-points in the Brillouin zones of the graphene sheets, generating twist-angle dependent saddle points in the band structures, and consequently effective gaps between the valence and conduction bands with corresponding  DOS van Hove singularities. The virtual crystal approximation-tight binding method also permits the determination of the optical absorption spectra for these tFLG systems. The calculated optical absorption spectra are found to be highly tunable via the twist angle. In particular, these spectra display clear peaks which correspond to the electronic  transitions from the saddle points in the valence band to those in the conduction band. The optical absorption results highlight the potential importance of the incommensurate tFLG systems for possible technological applications in graphene based tandem cells.

%%%%%%%%%%%%%%%%%%%%%%%%%%%%%%%%%%%%%%%%%%%%%%%%%%%

%%%%%%%%%%%%%%%%%%%%%%%%%%%%%%%%%%%%%%%%%%%%%%%%%%%
\end{document}